\documentclass[a4paper]{spie}

\usepackage[]{graphicx}
\usepackage{aas_macros}
\usepackage[]{subfigure}

\title{X-ray polarimetry on-board HXMT}

\author{Paolo Soffitta~\supit{a}, Ronaldo~Bellazzini\supit{b},
Gianpiero~Tagliaferri\supit{c},
Enrico~Costa\supit{a},\\Giovanni~Pareschi\supit{c},Stefano~Basso\supit{c},
Vincenzo~Cotroneo\supit{c},Massimo~Frutti\supit{a},\\Francesco~Lazzarotto\supit{a},
Fabio~Muleri\supit{a,d}, Alda~Rubini\supit{a}, Gloria~Spandre\supit{b},
Alessandro~Brez\supit{b}, Luca~Baldini\supit{b},
Jean~Bregeon\supit{b},\\Massimo~Minuti\supit{b},Giorgio~Matt\supit{e},
Filippo~Frontera\supit{f} \skiplinehalf \supit{a} Istituto di Astrofisica Spaziale
e Fisica Cosmica, Via del Fosso del Cavaliere 100, I-00133 Roma, Italy;
\\
\supit{b} Istituto Nazionale di Fisica Nucleare, Largo B.
Pontecorvo 3, I-56127 Pisa,  Italy
\\
\supit{c} Osservatorio Astronomico di Merate, Via E. Bianchi 46,
20121 Milano
\\
\supit{d} Universita' di Roma Tor Vergata, Dipartimento di Fisica, via della
Ricerca Scientifica 1, 00133 Roma, Italy
\\
\supit{e} Universita' degli Studi di Roma Tre,largo San Leonardo
Murialdo, 1, I-00146 Roma
\\
\supit{f} Universita' degli Studi di Ferrara, Via Savonarola 9,
I-44100 Ferrara, Italy}
\authorinfo{Further author information: (Send correspondence to Paolo Soffitta)
\\Paolo Soffitta: E-mail: Paolo.Soffitta@iasf-roma.inaf.it,
Telephone: +39-064993006}

\begin{document}
\maketitle

\begin{abstract}
The development of micropixel gas detectors, capable to image tracks produced in a
gas by photoelectrons, makes possible to perform polarimetry of X-ray celestial
sources in the focus of grazing incidence X-ray telescopes. HXMT is a mission by
the Chinese Space Agency aimed to survey the Hard X-ray Sky with Phoswich
detectors, by exploitation of the direct demodulation technique. Since a fraction
of the HXMT time will be spent on dedicated pointing of particular sources, it
could host, with moderate additional resources a pair of X-ray telescopes, each
with a photoelectric X-ray polarimeter (EXP$^{2}$, Efficient X-ray Photoelectric
Polarimeter) in the focal plane. We present the design of the telescopes and the
focal plane instrumentation and discuss the performance of this instrument to
detect the degree and angle of linear polarization of some representative sources.
Notwithstanding the limited resources, the proposed instrument can represent a
breakthrough in X-ray Polarimetry.
\end{abstract}


\section{Introduction}
Polarimetry is the sub-topic of X-Ray Astronomy for which we have the maximum gap
between the expectations deriving from theoretical analysis and the achievements
deriving from experiments. The main reasons for this are the limited sensitivities
in terms of  both statistics and control of systematics of  the conventional
techniques of Compton/Thomson scattering around 90$^{o}$ and Bragg diffraction at
45$^{o}$. The recent but impetuous development  of the micropixel polarimeters,
based on the visualization of photoelectron tracks in a gas, makes possible the
design of experiments with much higher sensitivity and an excellent control of
systematic effects. A sample of high luminosity and/or high expected degree of
polarization could now be studied even with telescopes of modest area: this
includes Crab Nebula/PSR, Accreting Binary, Pulsars, Black Hole Binaries and the
brightest AGNs. But polarimetry should be a tool capable to solve crucial problems
on several classes of X-ray emitting sources, provided that an adequate telescope
is available. Models predicts polarizations of the order of a few percent in low
luminosity sources, such as AGNs, or in sources showing a high luminosity in a
short time interval, such as millisecond binary pulsars or Soft Gamma Repeaters.
The detection of polarization in this range requires the detection of 10$^{5}$ -
10$^{6}$ photons, to be compared with a number of photons of the order of 10,
required to detect a source in an image, and this can be achieved only with an
adequate collecting area. The reverse of this limit is that polarimetry, when
performed in the focus of a telescope with an imaging detector, is in any case
performed on sources much brighter than the background in the image pixel. The
polarimetric sensitivity is only a matter of number of photons, independently
whether they are collected with a single telescope or with a cluster. This is not
true for imaging, where the sources at the limit of sensitivity are by definition,
comparable with background and the passage from one telescope to a cluster is
never for free. This opens the possibility to build up adequate collecting areas
by adding more than one telescope of shorter focal length to be harbored aboard a
bus much less demanding in terms of launcher or to be included as piggy-back
package on another mission.

Due to the room and weight limitations the inclusion of a focal
plane polarimeter aboard HXMT can be only based on the design and
manufacture of a dedicated telescope. With the proposed design and
with a combined use of the polarimeters and of the Hard X-ray
Telescope we set up a very performing mission,  much more than a
simple pathfinder.

In the following we describe the status of development of the Gas
Pixel Detector (GPD) and its polarimetric capabilities, and
propose a configuration of telescopes, that could be compliant
from the point of view of volume and weight with the launcher bay.
The derived sensitivities are very attractive.

In the paper we give for each subsystem the level of readiness of the technologies
involved and the strategies assumed to arrive to a B1 phase design all under
control or have very safe back-up solutions. None is based on items subject to
export restrictions out of control of the involved authorities.

\section{The Detector}
The X-ray polarimeter is basically a gas detector with fine 2-D position
resolution we developed (\cite{Costa2001}) to exploit the photoelectric effect.
Basically the polarimeter is made of a gas cell with an X-ray window, made of
Beryllium, a gas electron multiplier (GEM) which amplifies the electron tracks
generated in the drift gap and provides energy and time information and, below it,
of a pixellated plane which collects the charge content in each pixel to be
processed by the readout electronics.

\textbf{Principle of operation} Polarized X-rays introduces a 'cos$^{2}$'
asymmetry with respect to the azimuthal angle $\varphi$, the angle between the
direction of emission of the photoelectron  and the polarization vector (see for
example \cite{Compton1935}). The track produced by the photoelectron in a gas
mixture brings memory of the polarization of the original polarization vector.
Imaging the tracks with modern gas pixel detector permits to derive the absorption
point and the initial direction of the photoelectron path. From the distribution
of emission angles we derive the degree and polarization angle of the X-ray
photon. The Auger electron instead is not modulated by the X-ray polarization
therefore represent a disturbance, especially at low energy.

\textbf{The chip} The request of usage of the polarimeter in real astrophysical
experiment, in particular toward the reach of an area matched with the PSF and
Field of view of typical X-ray telescopes  has pushed us to leave the first
pioneeristic technology of a pixellated plane made with PCB multi-layer technology
and the electronics on a side. This technology was in fact limited either in the
number of pixel ($\sim$ 1000) and in the pixel pitch ($\sim$100 $\mu$m). We
therefore decided to adopt the solution to  bring the electronics inside the
detector gas cell by designing and building a multilayer ASIC CMOS analog chip
made with  0.18 $\mu$m VLSI technology. The top layer of the ASIC chip is
pixellated and is the readout plane of the charge produced in the gas volume and
amplified by the GEM. Two generations of chip are already behind us. The third
generation chip is currently working (\cite{Bellazzini 2006}) has the
characteristics reported in Table 1.


\begin{figure}[htbp]
\begin{center}
\subfigure[\label{fig:photoel}]{\includegraphics[angle=0,
totalheight=5cm]{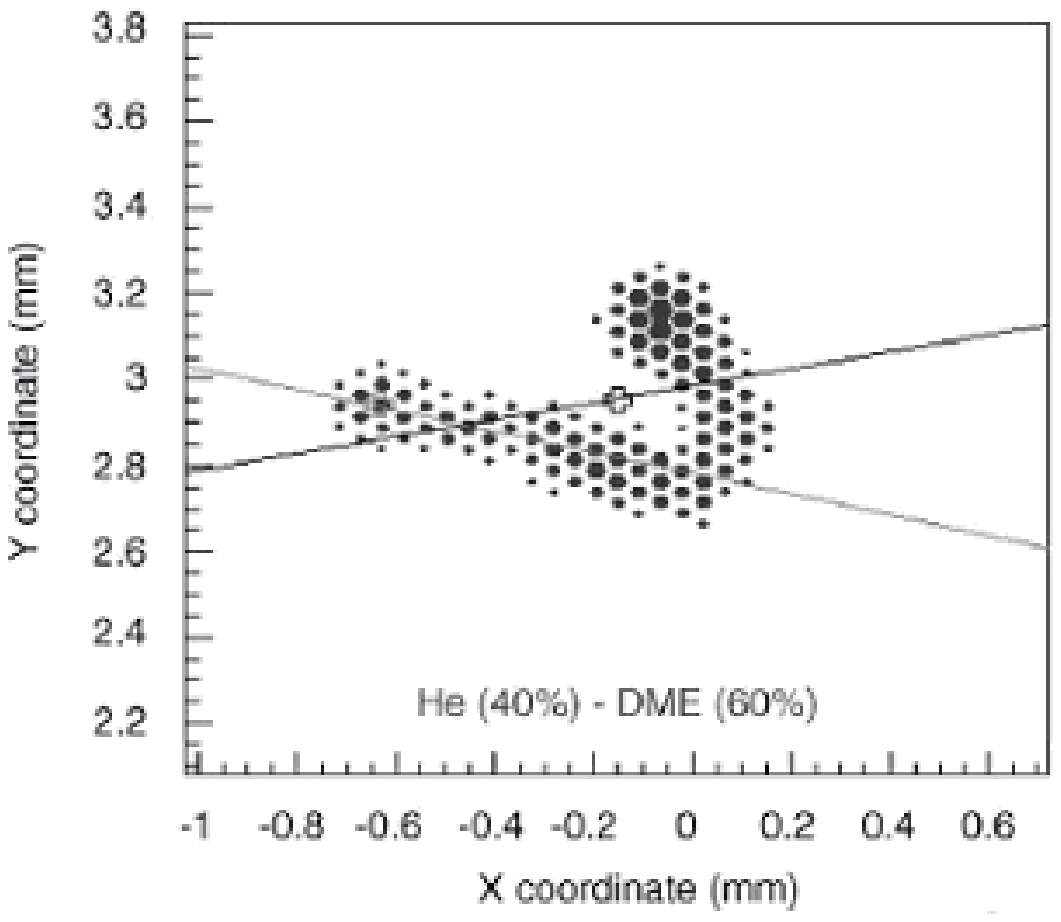}}
\subfigure[\label{fig:photoel_multi}]{\includegraphics[angle=0,
totalheight=5cm]{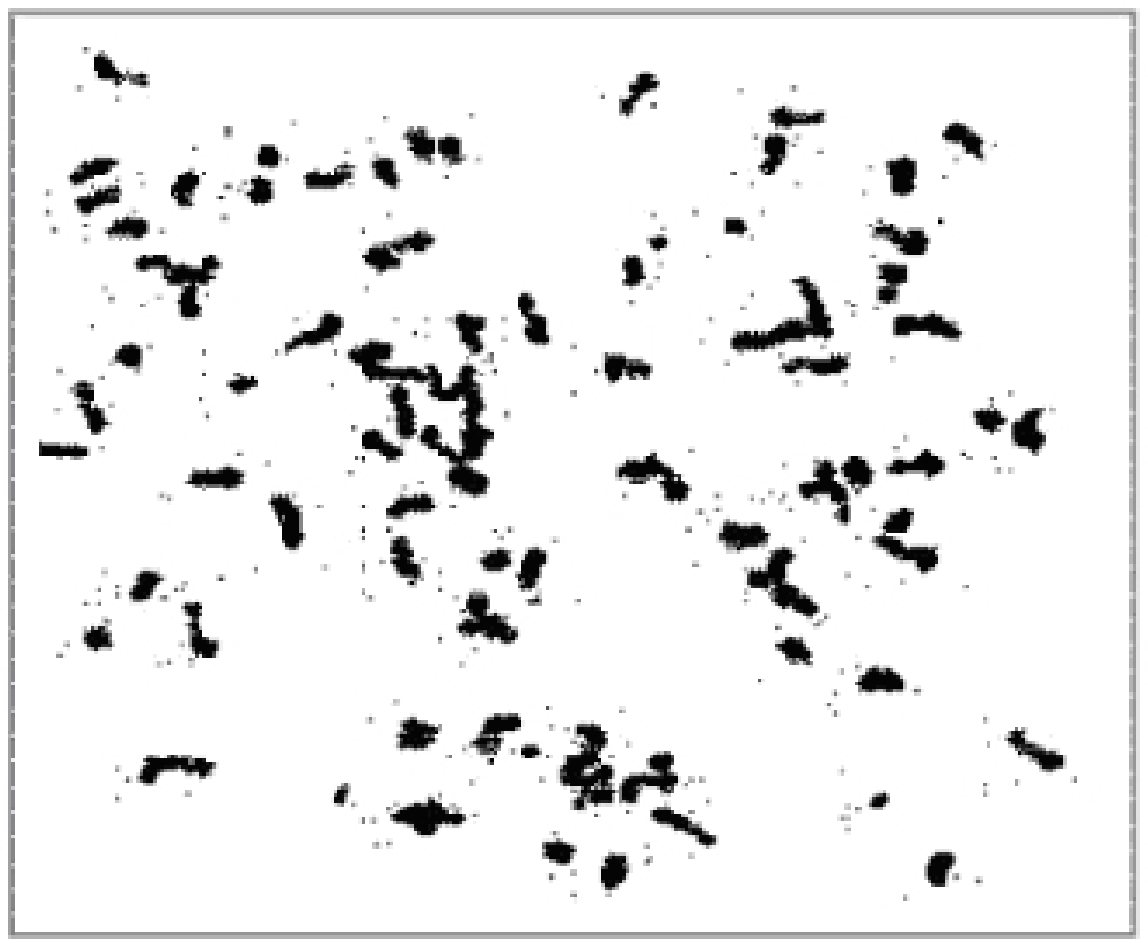}}
\end{center}
\caption{\small Tracks from the Gas Pixel Detectors. ({\bf a}) Single
Photoelectron Track. ({\bf b}) Collection of photoelectron tracks imaged by the
GPD in few seconds of data}
\end{figure}

\begin{table}[htbp]
\begin{center}
\begin{tabular}{c|c}
\hline \hline
Pixels & 105600 \\
Pitch & 50$\mu$m \\
Pixel Density & 470 Pixel/mm$^{2}$ \\
Area & 15 mm x 15 mm \\
Peaking time & 3-10 $\mu$s \\
Read-out & Asynch, Synch\\
Trigger & Intern, extern, self \\
Clock & 10 Mhz \\
Access to pixel & Direct,Serial,Full Matrix, ROI
\end{tabular}
\caption{Main Characteristics of the 105600 Pixel ASIC chip}
\label{tab:Crystals}
\end{center}
\end{table}

Each pixel is made of a metal pad, is connected to an individual charge-sensitive
amplifier and a shaping circuit.  The chip usually operates in a Region of
Interest mode (ROI) which downloads the charge content in each pixel within a
rectangular area which surrounds the pixels that passed the threshold, in order to
include all the photoelectron track  plus a selectable margin of 10 or 20 pixels.
The coordinates of the ROI are provided as soon as an internally triggered event
has terminated. The characteristics of the chip are already suitable for an
astrophysical experiment and we do not envisage great changes to be done but only
customization if any. In case of HXMT the performances of the chip are suitable
for the mission requirements and we are planning to use it as it is. In
~\ref{fig:photoel} and in ~\ref{fig:photoel_multi} we show a sample of a track
collected for a He-based mixture and a collection of 100 tracks imaged from the
same chip.

\textbf{The gas mixture}. The gas mixture which fills the polarimeter determines
the working energy band. To be effective, the energy of X-ray photons should be
sensitively larger than Auger electrons energy and thence of the k-shell energy of
the heaviest elements of the gas that dominate the absorption. Mixtures containing
Ne or He or organic molecules are preferred. Mixtures filled with Ar can be
effective only above 6 keV. Since the diffusion during the drift should be kept
small DME is best suited as quencher over Carbon-Dioxide or Methane. When in
combination with He, DME is the gas where most of the X-ray interaction occurs and
He is used as 'filler'. We studied different gas mixtures based with He and Ne and
DME as quencher. At the moment the mixture which provides the better performances,
as shown  below, are the He based mixture in the 2-10 keV energy band. We usually
worked at a gas pressure of 1-atm which is our baseline pressure. We use  a
mixture DME 80 \% He 20 \% as the baseline.

\begin{figure}[htbp]
\begin{center}
\includegraphics[angle=0, totalheight=6cm]{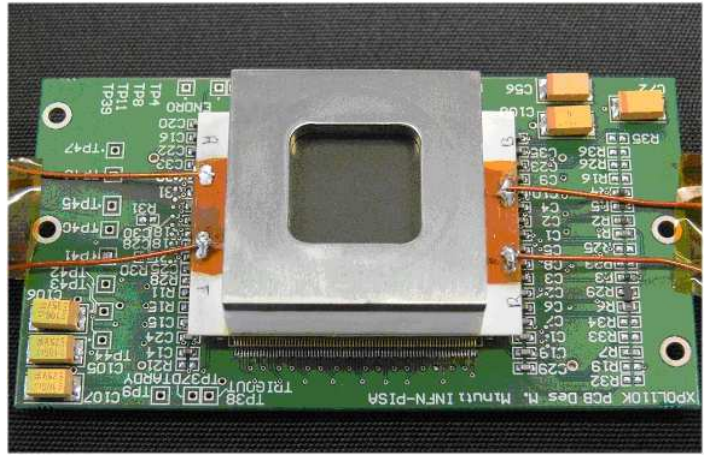}
\end{center}
\caption{The sealed gas pixel detector (GPD) \label{fig:GPD}}
\end{figure}

\textbf{The Body}. The detector body  hosts the window, the electrodes, the
feed-throughs and the readout chips with its package. The package brings-out 300
signals from the chip. The X-ray window provides, also, the drift field at -2000
V. The GEM  is a thin kapton foil metalized on both sides, perforated by
microscopic holes (50 $\mu$m pitch) with the top at -750 V and the bottom at -300
V. The chip is at ground. During the development phase we used detectors with
flowing gas system. We have built later, a sealed detector capable to withstand a
long term operation in space environment. It is manufactured  with materials and
components with low out-gassing rate. Ceramic component as Alumina for the
spacers,  kapton and ceramic feed-throughs are components of the body. The
distance between the chip and the window (the absorption/drift gap) is 1-cm.
Various prototypes of this sealed detector have been produced and extensively
tested along more than one year. The prototype detector is shown in
fig.~\ref{fig:GPD}.

\textbf{Stability vs time}. The detectors have shown no degradation of any
performance during an extensive testing lasting  months. One has undergone high
radiation testing for two months and further serendipitous testing for more than
one year. All time long it was irradiated with a rate higher than that foreseen
for the brightest sources in the focus of HXMT telescope.

\textbf{Disposition} The detector is directly mounted on a board. This is part of
a very compact Interface Electronics that performs the A/D conversion, the
suppression of  zeros, the tagging of events with time and forwards a packet of
data to a processor hosted within the Control Electronics. The detector and
Interface electronics are included in a box (10 cm $\times$ 10 cm $\times$ 5 cm).

\textbf{Performance and survival  in temperature}. Since the performance tests
have shown a certain sensitivity of the gain to temperature (as any other detector
based on gas multiplication) a Peltier cooler is included in the box to stabilize
the temperature of the detector around 10$^{o}$C. The detector and the Interface
electronics were thermal cycled and thermal-vacuum cycled between -15$^{o}$C and
45$^{o}$C showing beside survival, no degradation in the performances.

\textbf{Radiation Hardness}. The detector was irradiated with a Fe$^{55}$ source
to collect as much as radiation foreseen for the lifetime in orbit without loss of
performance. We foresee further testing with ions to improve the confidence on the
radiation hardness of the GEM.

\subsection{The focal plane} \label{sec:Source}
\begin{figure}[htbp]
\begin{center}
\includegraphics[angle=0, totalheight=6cm]{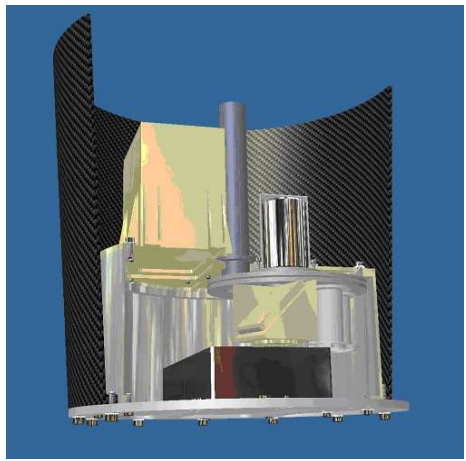}
\end{center}
\caption{Allocation of different parts in the focal plane of HXMT included the
polarized calibration source. \label{fig:Artistic_Draw}}
\end{figure}

In the fig~\ref{fig:Artistic_Draw} we report a sketch of the focal plane of the
polarimeter. It is composed by a box for the electronics (on the top left), where
the interface electronics can be placed, a filter wheel (on the right) and high
voltage power supplies to be located close to the detectors not shown in the
figure.

We envisage to use  a dedicated filter wheel. The use of the filter wheel procures
many advantages. Other than the 'Open' position the inclusion of a 'closed'
position permits to gather and monitor internal background rate along the orbit.
Two positions for two different low-rate calibration sources permit to check the
gain stability in-orbit and calibrate the gas-gain. One of the sources will emit
(unpolarized) fluorescence photons. The other source will radiate the detector
with photons polarized \cite{Muleri2007}by means of a Bragg diffraction around
45$^{o}$(shown in the fig.~\ref{fig:Artistic_Draw}).  A position with a diaphragm
permits to eliminate a possible bright source on the edge of the field of view.
The interface electronics will be located close to the detectors but connected via
a flexi-cable to have more freedom in the distribution of weights and thermal
inputs. To exploit at best the low energy optics bandpass, we envisage to use two
detectors, filled with slightly different mixtures. These could be interchanged in
the focal plane with a motorized linear stage, to optimized the measurements on
the observed source. The development cost ad the resources needed for the
alternate operation of the detectors are very limited, since the detectors are
just filled with different mixtures and the control electronics could be shared.
Moreover the functionality of the filter wheel and of the linear stage could be
joined in an unique motorized stage, reducing the total weight of the focal plane.
We are currently studying the scientific improvements achievable by means of the
inclusion of a polarimeter softer than the current He-DME baseline.

\section{The interface electronics and the control electronics} \label{par:MechanicalAssembly}

The Interface Electronics (IE) resides close to the detector body eventually below
it. Mainly consists of an  ADC which converts the analog signals from the chip and
stores them in the DPRAM. An FPGA  handles and generates the signals for the chip
and routes the data coming from a DPRAM  to the control electronics.  An  actual
board of interface electronics is currently working in laboratory. The Control
Electronics (CE) contains the Data Processing Unit (21020 DSP as a baseline) a
board which handles the telecommands and a board with DC-DC converters which
provides the supply to the chip. The mass-memory shown in the figure is also
included in the Control Electronics. We are open to discuss an onboard mass memory
shared with HXMT payload or with the bus but as a baseline we prefer to assume
that EXP$^{2}$  has its own mass memory.

\section{ The telescopes}
In our baseline we assume a coating for the outern surfaces of shells of Iridium
with a thin over-coating of carbon. The Iridium, due to its higher density
(resulting in larger critical angles), provides a better reflectivity than gold at
higher energies. A thin carbon layer "fills" the reflectivity decrease around the
M absorption edges, that for Au and Ir is around 3 keV where the polarimeter is
particularly sensitive. The deposition of Iridium is different from that of Gold.
It has been tested in several contexts and we do not foresee any major problem to
introduce it in the process of manufacture of the mirrors. For soft X-ray
telescopes based on the Ni replication technology operated so far (SAX, XMM,
SWIFT�) it was used an Au reflecting coating, because gold also acts as the
release agent in the replication process due to its low adherence onto the
superpolished mandrel surface and good adherence to the nickel. On the contrary,
both Iridium and Carbon present a strong adherence to the mandrels and, then, have
to be applied to the gold surface in a post-process coating deposition after
having followed the usual replication method. The deposition of carbon is also a
very well established operation. The application to HXMT mirrors could be
difficult for the inner mirrors but should be feasible for outer mirrors where
they give the dominant contribution at low energies. Anyway the adoption of this
coating does not interfere with the manufacture of mandrels which is the most
demanding part of the process. In case some unexpected obstacle is found the
traditional coating with gold is a very safe back-up solution. In this worst case
the effective areas given below must be scaled down by about a 20\%.

Two identical telescopes with the same focal length of 2100 mm are foreseen. Each
telescope is composed by 30 shell with diameter ranging from 90 to 270 mm. Iridium
and Carbon coating is foreseen for the 22 outer shells.

The manufacture of the telescope HXMT has an ambitious time schedule, that
discourages the application of technologies requiring a too long and uncertain
development phase. The volume available for the X-ray telescope is filled with a
reasonable efficiency and with an acceptable weight with telescopes manufactured
with the technology of replicas of superpolished mandrels by Ni electroforming of
the mirror shell walls. Moreover this is completely under control of Italian
institutes and industries, and a design/manufacture/test planning can be set up
under the control of Italian and Chinese Space Agencies, without depending on
third subjects. For this reason  we propose here solutions based on this
technology. The design we propose is therefore based on this technique.The
thickness of the shells is fixed to 0.2 mm. The possibility to reduce this
hickness to 0.1 mm at least for a number of shells is under study. This is not
necessarily the best solution but allows for a very realistic evaluation of the
weights of the shells and of the mechanical structure needed to keep them mounted
and aligned. This means that the proposed designs are feasible with the declared
weights with high level of confidence. This will result in an optical quality of
the order of 30", ~Half Energy Width (HEW), that, combined with the effects of
inclined penetration in the gas will guarantee the resolution less than arcminute,
compliant with all the targets of the polarimeter for HXMT.

\begin{figure}[htbp]
\begin{center}
\includegraphics[angle=0, totalheight=14cm]{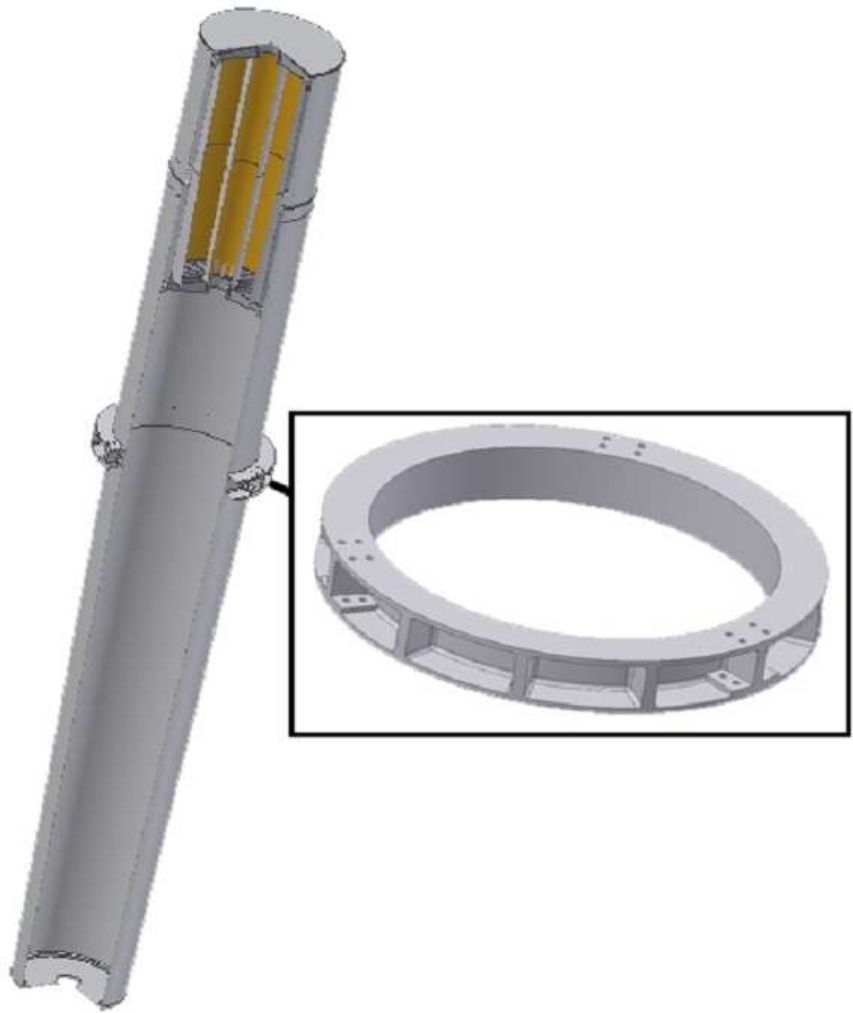}
\end{center}
\caption{The telescope tube which holds the mirror for HXMT.
\label{fig:telescope_Art2}}
\end{figure}

The design proposed is based on the well established technique of SAX, XMM and
SWIFT. The capability of producing shells thinner with a controlled loss of
angular resolution has been already tested by the production both of  a thin shell
from a mandrel of JET-X and of thin shells from mandrels developed as part of the
Simbol-X project. All these thin shells have been tested at the Panter Facility
providing HEW $<$ 30". To avoid any uncertainty in the manufacture integration and
testing of the flight units, two HXMT prototype shells will be produced and tested
in the connecting phase. Also the procedures for the Iridium and Carbon coating
will be assessed during the connecting phase. In case a complete confidence is not
achieved the Ir coating can be released and the C coating can be deposited via a
collaboration with other institutions (USA) that have a good control of this
technique and with which we have already a scientific collaboration in place (CfA)
for other projects.

\section{The integration of the polarimeter with HXMT}
To fully employ the free space in the bay of the launcher, we envisage to use two
identical "telescope units", each composed by a telescope and by the focal plane
instrumentations with different power supplies, filter wheels and Interface Front
End Electronics, enclosed in a thin carbon fiber tube. We assume that the mirrors
are positioned with the so-called spiders within the tube structure which is then
positioned on a bulk-head to be fixed to a side of HXMT see
fig.~\ref{fig:telescope_Art2}. The tube is connected to the spacecraft through a
flange. The tube itself is divided in two part, both mounted on the interface
flange. A forward tube on which is mounted the mirror module and that carries also
the closing door on top, and a tapered rear-tube that will hold the focal plane
detector. The flange will divide the tube at the center of mass.

Since the baseline configuration of the detector is effective above 2 keV, we can
afford a thin thermal blank in front of the mirrors and we do not foresee a long
baffle as in the case of JET-X. The thermal stabilization of the mirrors is much
less demanding (of the order of 15 W for both telescopes). We foresee a protective
cover in front of each telescope to be opened when in orbit. A design of a
possible interface of the telescopes with the bus can be seen in
fig.~\ref{fig:telescope_Art2}, in practice we expect to fix the flange to the bus
with a three points connection.

\section{Expected performances}
\textbf{Space/angular Resolution} The reconstruction of the impact point of the
photon can be performed with a resolution (FWHM) of 150 $\mu$m for an orthogonal
beam. In the focus of a telescope the photons are impinging with a certain
inclination and will be absorbed at an unknown depth in the gas. This introduces a
further, and prevailing, uncertainty in determining the position where the photon
has crossed the focal plane. This uncertainty is of the order of 400 $\mu$m. The
angular resolution is the combination of the position resolution, divided by the
focal length, and the resolution of the optics. Assuming that the errors combine
with a quadratic law we find that our system will have a point spread function of
one arcminute diameter.

\textbf{Energy Resolution} The MPGD has an energy resolution of the order of that
of a Proportional Counter. We can assume: (DE/E)FWHM = 0.2 � (6/E(kev))$^{1/2}$
The detector could, in principle, perform better than this but it must be verified
that this is consistent (e.g. in terms of gain) with the optimal use as a
polarimeter.

\textbf{Timing Capability} Events  can be tagged with the arrival time with a
resolution of few $\mu$s. We fix a resolution to 8$\mu$s on the basis of
astrophysical considerations.

\textbf{Polarimetric capability} Polarimetric capability depends on the filling
gas mixture that is to be tuned to the optics band-pass and to a trade-off of
different astrophysical targets. In the case of a multi-detector design the
filling gas would be different for each detector. We use as a reference a filling
with 1 atmosphere of a mixture 20 \% He 80 \% DME (Dimethyl Ether), which is
presently giving very interesting results, confirmed by recent measurements
\cite{Muleri2008}, but is not necessarily the final choice. For instance a larger
fraction of He (e.g. 30 \%) provides better results at low energies. Since the
sensitivity of a polarimeter is proportional to $\epsilon$$^{1/2}$$\times$$\mu$,
where $\epsilon$ is the efficiency and $\mu$ is the modulation factor, we use this
quantity for the above mentioned mixture. In practical cases this has to be
convolved with the effective area of the optics and with the spectrum of the
source, and integrated in finite intervals of energy. This quantity gives an idea
of the energy at which the device is performing better. The best sensitivity is
reached at about 3 keV.

\section{Astrophysical performances}

\begin{figure}[htbp]
\begin{center}
\includegraphics[angle=90, totalheight=10cm]{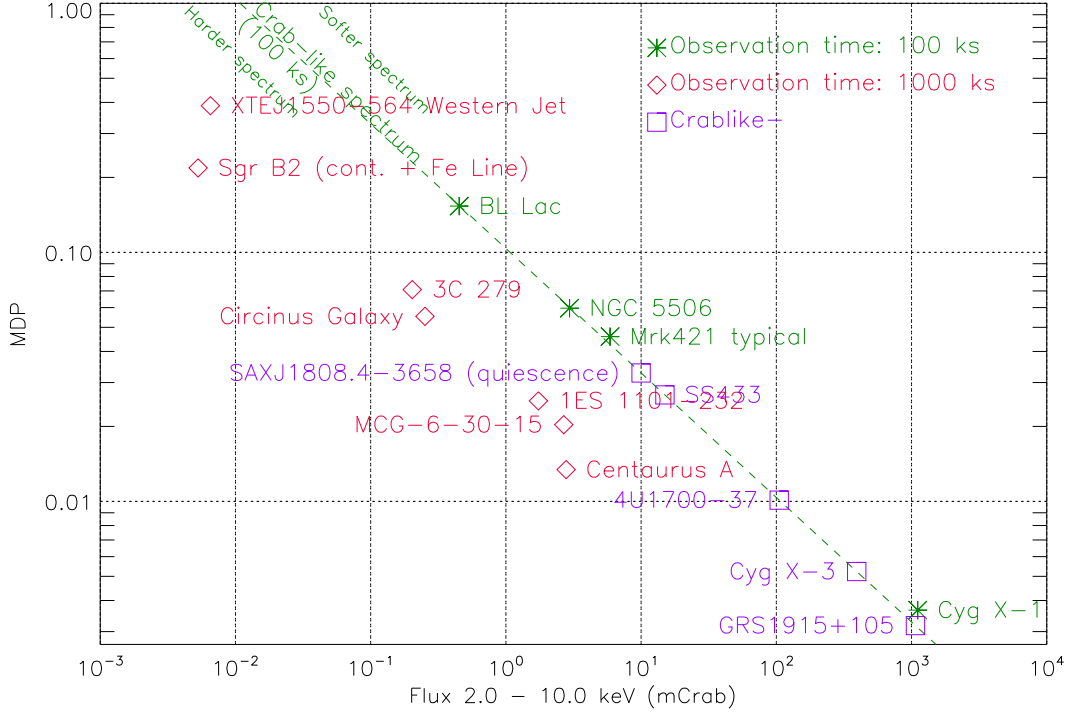}
\end{center}
\caption{The minimum detectable polarization for a selection of sources.
\label{fig:MDP}}
\end{figure}

By combining the effective area of the telescopes with the polarimetric
capabilities of EXP$^{2}$,  we can evaluate the astrophysical performances of the
experiment. The Minimum Detectable Polarization with a 3$\sigma$ confident level
for the current baseline configuration is reported in fig.~\ref{fig:MDP} in the
energy range 2 - 10 keV and for a selected sample of sources.

The background is evaluated considering both the internal background from low Z
(neon or methane) proportional counter flown on board of OSO-8 and the diffuse
X-ray background. The internal background is 1.5 10$^{-4}$ counts/s/cm$^{2}$/keV
(1.6-3 keV ) and 1.0 10$^{-4}$ counts/s/keV/cm$^{2}$ (3-10 keV). We expect a total
internal background of (c/s) =  1.0 10$^{-5}$ and a total diffuse background of
(c/s) = 1.0 10$^{-8}$. The first of these two figures could be underestimated, but
in any case our system will be source-dominated for orders of magnitude. This
assures that the assumption of adding the two telescopes efficiencies to determine
the total sensitivity is correct. X-ray polarimetry will be performed on all the
class of sources of X-ray astronomy except, may be, for cluster of galaxies. A
goal well within the reach is the observation of High Mass X-ray binaries. For the
data of X-ray pulsators we can assume that even performing the polarimetry on
selected phases (namely splitting the observing time) we are still sensitive to a
few \% level. Therefore EXP$^{2}$ can measure the swing of the polarization angle
with phase, discriminate between fan beam and pencil beam and fix all the geometry
of the system (\cite{Meszaros1988}).

\begin{figure}[htbp]
\begin{center}
\subfigure[\label{fig:AoP}]{\includegraphics[angle=90, totalheight=5cm]{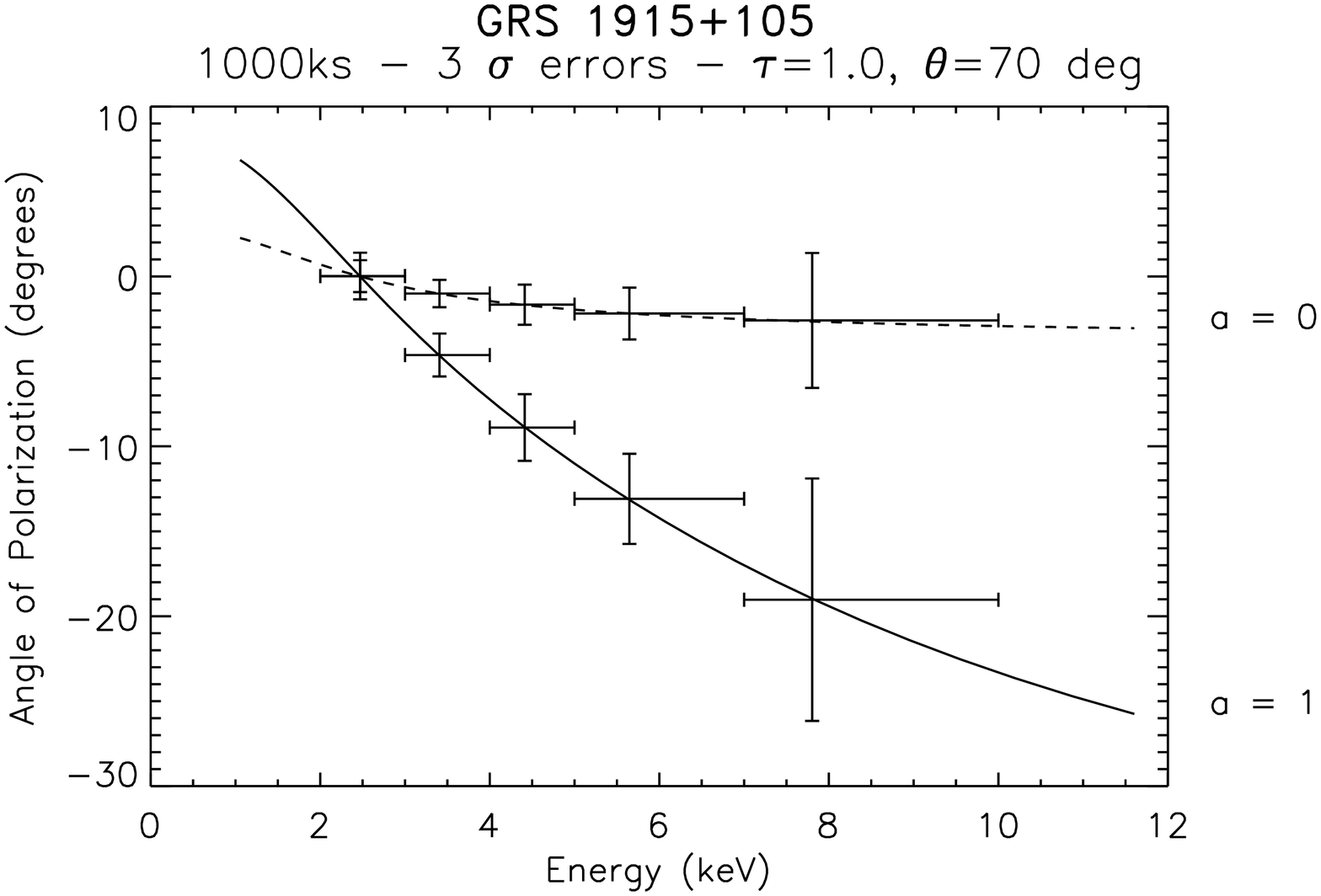}}
\subfigure[\label{fig:DoP}]{\includegraphics[angle=90, totalheight=5cm]{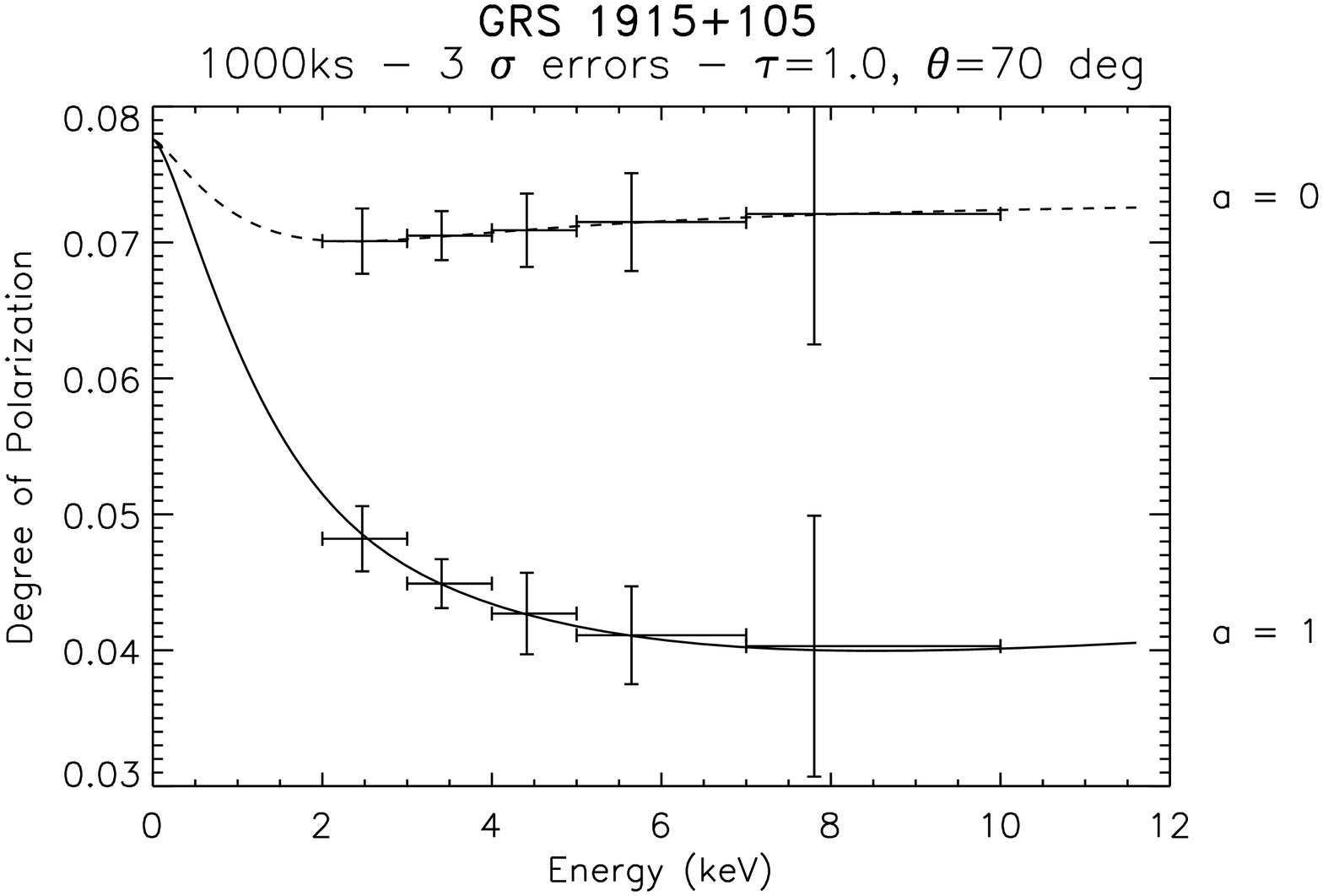}}
\end{center}
\caption{\small Polarization properties simulated from GRS1915+105. ({\bf a})
Expected rotation of the polarization angle with energy. ({\bf b}) Expected
variation of the degree of polarization with energy.}
\end{figure}

\begin{figure}[htbp]
\begin{center}
\includegraphics[angle=0, totalheight=7cm]{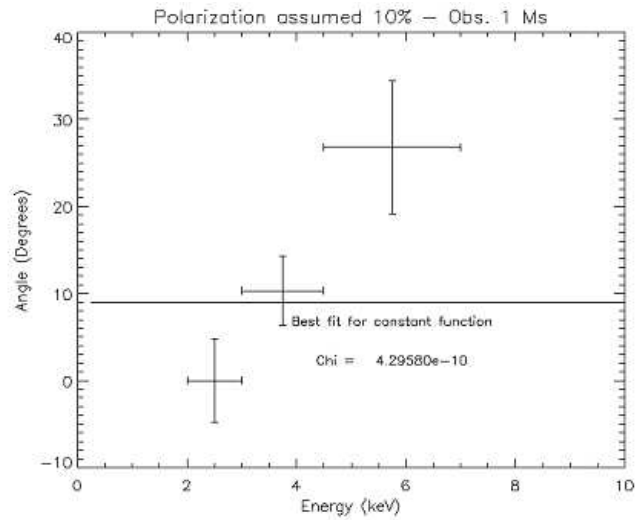}
\end{center}
\caption{1Ms  observation of the blazar 1E 1101-232 (z=0.186)  allows to reach an
upper limit in the vacuum birifrangence in the Quantum Loop Gravity with
unprecedent sensibility. \label{fig:QG}}
\end{figure}
The polarimeter on-board HXMT will be able to perform significative measurements
on board the most luminous AGNs.  From the sensitivity to Blazar we can be
confident that EXP$^{2}$ can confirm the role of synchrotron on the emission
process. For blazar with comptonized spectrum it will be possible to discriminate
whether the seed photons come from the Broad Line Regions or from the Disk or are
due to Self Synchrotron from the jet. Even looking to the galactic center
EXP$^{2}$ will disentangle the origin of the emission of molecular clouds. A long
pointing of Sgr B2 will allow for an MDP of 16 \%. This can (and must) be expected
in the reflection model. Moreover the angle should "point" to SgrA* demonstrating
that a few centuries ago our Galaxy was a low luminosity AGN. Coming to galactic
Black Holes and in particular micro-quasars polarimetry of soft-state when the
emission is derived mainly from the disk can be decisive in detecting General
Relativity effects. From the sensitivity of GRS1915+105 we have simulated
10$^{6}$s of observing time (see fig.~\ref{fig:AoP} and ~\ref{fig:DoP}) splitting
polarimetry on selected phases (namely splitting the observing time). We are still
sensitive to \% level and we can discriminate between Kerr and Schwartschild black
hole (\cite{Dovciak2008}. The quoted errors are at 3$ \sigma$. This shows that the
polarimeter aboard HXMT will be a breakthrough for many topics of High Energy
Astrophysics.

To these "guaranteed" results we can add a large space for discovery in particular
for the polarization by shocks in shell-like Supernova Remnants, moderately space
resolved polarimetry of the Crab Nebula (the prototype accelerator), polarimetry
of magnetars in outburst (a test of QED effects of vacuum birefringence and/or
proton cyclotron), polarimetry of quasars by scattering on the disc, polarimetry
of GRB afterglows. A further very attractive possibility is the test of Loop
Quantum Gravity models by searching (see fig.~\ref{fig:QG}) for the rotation of
the polarization angle with energy and distance in Blazars \ref{fig:QG}. We
simulated the effect of Quantum Gravity on a fairly distant blazar to show at
which level the linear term can be excluded\cite{Gambini1999}.

\section*{Acknowledgments}
This work is supported by a contract of Italian space Agency

\bibliography{References}   
\bibliographystyle{spiebib}   

\end{document}